\documentclass[journal,twoside,web]{ieeecolor}
\usepackage{lcsys}
\usepackage{cite}
\usepackage{amsmath,amssymb,amsfonts}
\usepackage{algorithmic}
\usepackage{graphicx}
\usepackage{textcomp}
\usepackage{epstopdf}
\usepackage{setspace}
\usepackage{enumerate}
\usepackage{stfloats}
\usepackage{graphics} 
\usepackage{subfigure}
\usepackage{epsfig}
\usepackage{color}

\allowdisplaybreaks 
\allowdisplaybreaks[4]

\usepackage{lettrine}
\usepackage{indentfirst}
\usepackage{CJK}
\usepackage{multicol}
\usepackage{float}
\usepackage{setspace}
\usepackage{amsfonts}
\usepackage{caption}
\usepackage{upgreek}
\usepackage{mathtools}
\usepackage{etoolbox}
\usepackage[colorlinks=true, linkcolor=black, citecolor=black, urlcolor=black]{hyperref}
\newtheorem{remark}{Remark}
\newtheorem{assumption}{Assumption}

\newtheorem{theorem}{Theorem}

\newtheorem{lemma}{Lemma}
\newtheorem{definition}{Definition}
\newenvironment{ProofNoIndent}[1][Proof]{%
  \par\vspace{0em}%
  \indent\textit{#1:}\hspace{0.5em}%
}{\hfill$\blacksquare$\par}

\def\BibTeX{{\rm B\kern-.05em{\sc i\kern-.025em b}\kern-.08em
    T\kern-.1667em\lower.7ex\hbox{E}\kern-.125emX}}
\begin{document}
\title{Extended Version of ``Distributed Adaptive Resilient Consensus Control for Uncertain
Nonlinear Multiagent Systems Against Deception Attacks''}
\author{Mengze Yu, Wei Wang, \IEEEmembership{ Senior Member, IEEE}, and Jiaqi Yan
\thanks{This is an extended version of a paper accepted to IEEE Control Systems Letters. }
\thanks{This work was supported by the National
Natural Science Foundation of China under Grant 62373019. \textit{(Corresponding author: Jiaqi Yan.)} }
\thanks{Mengze Yu is with the School of Automation Science and Electrical Engineering, Beihang University, Beijing 100191, China (e-mail: yumengze0214@126.com).}
\thanks{Wei Wang is with the School of Automation Science and Electrical Engineering, Beihang University, Beijing 100191, China, also with the Hangzhou Innovation Institute, Beihang University, Hangzhou 310051, China, and also with the Zhongguancun Laboratory, Beijing 100191, China (e-mail: w.wang@buaa.edu.cn).}
\thanks{Jiaqi Yan is with the School of Automation Science and Electrical Engineering, Beihang University, Beijing 100191, China, also with the Hangzhou Innovation Institute, Beihang University, Hangzhou 310051, China (e-mail: jqyan@buaa.edu.cn).}}

\maketitle
\thispagestyle{empty}
\begin{abstract}
This paper studies distributed resilient consensus problem for a class of uncertain nonlinear multiagent
systems susceptible to deception attacks. The attacks invade
both sensor and actuator channels of each agent. A specific
class of Nussbaum functions is adopted to manage the attack-incurred multiple unknown control directions. Additionally, a
general form of these Nussbaum functions is provided, which
helps to ease the degeneration of output performance
caused by Nussbaum gains. Then, by introducing finite-time
distributed reference systems and local-error-based dynamic
gains, we propose a novel distributed adaptive backstepping-based resilient consensus control strategy. We prove that all
the closed-loop signals are uniformly bounded under attacks,
and output consensus errors converge in finite time to a clearly-
defined residual set whose size can be reduced by tuning control
parameters, which is superior to existing results. Simulation
results display the effectiveness of the proposed controllers.
\end{abstract}


\section{Introduction}
\label{sec:introduction}
\IEEEPARstart{D}{istributed} consensus of multiagent systems (MASs) has drawn abundant attention in recent years and many representative results have been reported \cite{revier1,revier2,revier3}. It aims to realize an agreement on the state or output of each agent through local communication and control law, which widely possesses a range of potential applications in plenty of military and civilian fields \cite{3,4}. However, MASs involve numerous communication channels and public protocols, making them vulnerable to data tampering by attackers, which can compromise these traditional control schemes. Hence, the concept of resilient consensus has been presented and has earned extensive attention in the control field. It aim to ensure that MASs can reach an acceptable agreement and remain stable despite the presence of cyber attacks.

So far, two categories of cyber attacks have earned considerable concerns, including denial-of-service (DoS) attacks \cite{5,6} and deception attacks \cite{7,8,9,10,11,12}. DoS attacks disrupt communication channels and prevent transmitted data from being received. Then, deception attacks hijack the data transmitted between sensor and  actuator and re-inject elaborate deceptive signals into the channels, impacting the consensus and stability of MASs. Generally, the attack-injected data may affect the operation of MASs from two aspects. Sensor injected data may incur intractable time-varying uncertainties in system dynamics, which compromises consensus control schemes that rely on accurate system models, potentially leading to their failure. In addition, the intrinsic control direction of each agent may be secretly reversed by actuator attacks. The study in \cite{wireless} shows that digitizing control signals for wireless transmission introduces such vulnerabilities. Specifically, the control inputs $u$ are typically encoded as signals $d(u)$. An attacker may manipulate the modulation by reversing the symbol of $d(u)$, driving the actuator to decode an inverted command $-u$, severely ruining the closed-loop stability. Hence, it is vital to develop resilient controllers to resist deception attacks on the MASs and maintain admissible system performance.

Over the past few decades, several control techniques have been widely studied to solve the aforementioned problems. The first is adaptive control, which renders conspicuous effectiveness in relieving the impact of multifarious uncertainties \cite{revier3,3,4}. Besides, the Nussbaum function technique is widely used to identify and track the true control direction of controlled plants \cite{13,14,15}. Hence, there has reported a list of remarkable researches based on these techniques to ease the malicious impacts of deception attacks and achieve resilient consensus \cite{7,8,9,10,11}. Early attempts are mostly oriented toward linear agents \cite{7}, whereas nonlinear phenomena are universal in practical physical systems. Backstepping controllers are well-studied and effective for upper-triangular nonlinear systems \cite{revier3,3,4}, leading to various resilient consensus protocols for such high-order systems. In \cite{9,10,11}, adaptive parameters are drawn to mitigate attack-induced uncertainties, distributed controllers are designed to ensure the boundedness or asymptotic convergence of consensus errors. In particular, \cite{11} uses enhanced Nussbaum functions called $\mathcal{N}$\emph{-Function} \cite{14,15} to address the problem that attacks render the actual control coefficients in each step inaccessible and nonidentical.

Note that some limitations still exist. As claimed in \cite{14}, $\mathcal{N}$\emph{-Function} plays a crucial role to handle  multiple unknown control directions. However, only a special class of functions $\mathcal{N}(\nu)=e^{a\nu^2}\sin(b\nu+c\pi/2)$ has been certified as $\mathcal{N}$\emph{-Function} in \cite{15}. Since the gain ($\nu(t)$) typically increases monotonically, the rapid growth of $e^{a\nu^2}$ may induce oscillations or spikes in system outputs, leading to poor dynamic performance. In addition, in \cite{10}, unknown integral terms related to $\kappa(t)$ are involved in the upper bound of the closed-loop Lyapunov function, leading to an unknown upper bound of the consensus error. How control parameters affect the consensus error also becomes vague. Moreover, \cite{9} and \cite{11} employ sign-approximation functions with exponential terms to handle attack uncertainties and achieve asymptotic consensus. However, as a cost, they may not strictly ensure the boundedness of all closed-loop signals over the entire time horizon.

Encouraged by these observations, we establish a new backstepping-based distributed resilient control strategy to achieve resilient consensus for a class of MASs manipulated by local-channel deception attacks. The main contributions are summarized as follows: (1) Since attacks on different states may be inconsistent, control directions of the first virtual controller and the control input in the backstepping design may be unknowingly differed.To handle this impact, $\mathcal{N}$-functions are wielded in controller design to achieve resilient consensus. Compared with \cite{15}, we further certify two more general forms of $\mathcal{N}$-functions, allowing for slower-growing functions and conducing to mitigate potential performance degradation. (2) Compared with existing methods like \cite{10}, the proposed scheme uses adaptive gains to regulate the first backstepping error of each agent, ensuring finite-time convergence of consensus errors to a clearly-formulated set. Moreover, the relationship between the error bound and controller parameters is also clarified. (3) Unlike resilient control results \cite{9,11}, which may not strictly ensure the stability of all closed-loop signals over the entire time horizon, the proposed controller ensures not only the convergence of output signals but also the overall closed-loop stability of the entire MASs.

The rest of this paper is constructed as follows. Section \ref{sec:ProblemFormulation} models the MASs and attacks and states the control objective. Section \ref{sec:DesignofDistributedAdaptiveResilientControllers} proposes a distributed resilient control framework with stability analysis. Simulation results are laid in Section \ref{sec:Simulation}, followed by the conclusion in Section \ref{sec:Conclusion}.

{\emph{Notations}: $\mathbb{R}^p$ stands for the set of real column vector with $p$ dimension. The set of positive integers is denoted by $\mathbb{N}^+$. $||\cdot||$ represents 2-norm of a vector. $\mathrm{diag}\{\cdot\}$ stands for diagonal matrices. Let $\lambda_2(\mathcal{Q})$ be the second smallest eigenvalue of matrix $\mathcal{Q}$.
Besides, the interval of time $t$ is set as $t\in [0, +\infty)$.

\section{Problem Formulation}
\label{sec:ProblemFormulation}
\subsection{System Model}

Here we focus on a group of multiagent systems with $N$ nonlinear agents as follows.
\begin{equation}
   \begin{aligned}
 \label{system model}
\dot{x}_{i,1}(t)&=\psi_{i,1}^T(x_{i,1})\vartheta_{i,1}+g_{i,1}(t)x_{i,2}(t)+o_{i,1}(t),  \\
\dot{x}_{i,2}(t)&=\psi_{i,2}^T(x_{i})\vartheta_{i,2}+g_{i,2}(t)\check{u}_i(t)(t)+o_{i,2}(t),\\
  y_i(t)&=x_{i,1}(t),
   \end{aligned}
   \end{equation}
where $x_{i}(t)\!\!=\!\!\![x_{i,1}(t),x_{i,2}(t)]^T \!\!\!\in \!\!\mathbb{R}^2$ and $y_i(t)$ denote the local system states and output of agent $i$ for $i = 1,\ldots, N$, respectively. Then, $\check{u}_i(t)\! \in \!\mathbb{R}$ refers to the actually-actuated input of agent $i$, $\psi_{i,j}(\cdot)\!\!\! \in \!\! \mathbb{R}^{v_{ij}}(v_{ij}\!\!\!\in \!\!\mathbb{N}^+, j=\!1,2) $ represent known nonlinear smooth functions multiplied by unknown parameters $\vartheta_{i,j}$, and $g_{i,j}(t)$ are piecewise-continuous non-zero time-varying functions. Especially,  $g_{i,2}(t)$ is denominated as the control coefficient. The piecewise-continuous disturbance is denoted by $o_{i,j}(t)$. Extensive studies \cite{revier3,9} show model \eqref{system model} outlines the dynamics of various engineering systems ($e.g.$, mechanical systems).  Note that the proposed framework can be extended to higher-order strict-feedback systems like \eqref{system model}.

\begin{figure}[t]
\setlength{\belowcaptionskip}{-0.5cm}
\centering
\includegraphics[width=6cm,height=2.3cm]{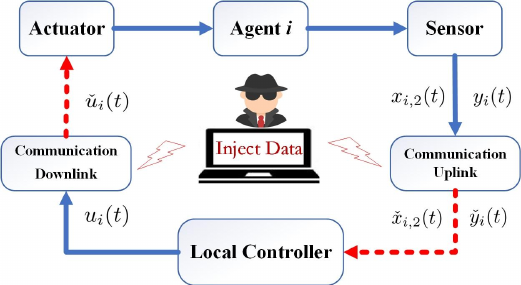} 
\caption{The behaviors of deception attacks.}
\label{System block}
\end{figure}

\subsection{Network Topology}
The communication topology of the MASs is depicted by a \emph{directed} graph $\mathcal{G}\triangleq (\mathcal{V},\varepsilon,\mathcal{A})$, $\mathcal{V}=\{1,\ldots,N\}$ is the set of indexes, $\varepsilon \subseteq \mathcal{V}\times\mathcal{V}$ is  the edge set, $\mathcal{A}=[a_{ij}]_{N\times N}$ refers to the adjacency matrix representing the non-negative weights of the communication. The edge $(j,i)\in \varepsilon$ means that agent $i$ can receive signals from agent $j$, let $\!a_{ij}\!\!=\!\!1$ and $j$ is a neighbor of $i$, otherwise $a_{ij}\!=\!$. Let $a_{ii}\!=\!0$. The set of agent $i'$s neighbors is $\mathbf{N}_i \!\triangleq \!\{j\in\! \mathcal{V}:(j,i)\!\in \!\varepsilon\}$. $\mathcal{D} = \mathrm{diag}\{d_1,... ,d_N\}$ is the indegree matrix with $d_i = \sum_{j\in \mathbf{N}_i}a_{ij}$. Laplacian matrix is $\mathcal{L} = \mathcal{D}-\mathcal{A}$.  A directed path from agent $i$ to $j$ is a sequence of successive edges $\{(i,k),(k,l),\ldots,(m,n),(n,j)\}$, where $k,l,\ldots,m,n$ are intermediate agents. In this paper, we consider that $\mathcal{G}$ is \emph{strongly connected}, $i.e.$, there exists a directed path between any two agents. Then, the following lemma is given.

\begin{lemma}[\!\!\cite{16}]\label{Lemma 1}
 For a \emph{strongly connected} graph $\mathcal{G}$, 0 is a simple eigenvalue of $\mathcal{L}$ with an eigenvector $\mathbf{1}=[1,\ldots,1]^T$. A positive left eigenvector $h\triangleq [h_1,\ldots, h_N]^T$ of $\mathcal{L}$ renders that $h^T\mathcal{L}=0$ and $h^T\mathbf{1}=1$. Define $ \mathcal{H}\triangleq \mathrm{diag}\{h_1,\ldots, h_N\}$ and $\mathcal{Q}\triangleq \mathcal{H}\mathcal{L}+\mathcal{L}^T\mathcal{H}$, $\lambda_2(\mathcal{Q})>0$. If $l \in \mathbb{R}^N$ is a positive vector, it follows that $x^T\mathcal{Q}x \!\geq \! \lambda_2(\mathcal{Q})N^{-1}x^T\!x, \forall l^Tx=0, x\in \mathbb{R}^N$.
 \end{lemma}
 \begin{remark}\label{remark topology}
Strong connectivity is common in leaderless consensus studies \cite{9,10,11}. It ensures bidirectional information flow among all agents, thereby enabling consensus. In contrast, weak connectivity allows only unidirectional transmission, which hinders full information exchange and is not conducive to attaining a consensus. Such topologies are more applicable for leader-following scenarios \cite{13}.
\end{remark}
 \vspace{-1mm}
\subsection{Attack Behavior}
We consider the case where the \emph{internal uplink and downlink} of \emph{each agent} are liable to be manipulated by deceptive adversaries. The behavior of these attacks can be depicted as
\begin{align}
\check{y}_i(t)&=\varrho_{o}(t)y_{i}(t), \label{2} \\
\check{x}_{i,2}(t)&=\varrho_{si}(t)x_{i,2}(t),\label{3}\\
\check{u}_i(t)&=\varrho_{ai}(t)u_i(t) \label{4}.
\end{align}
\indent As viewed in Fig. \ref{System block} and \eqref{2}-\eqref{4}, the sensor-measured output and state $y_{i}(t)$,\! $x_{i,2}(t)$ may be multiplicatively corroded by unknown injected attack weights $\varrho_{o}(t)$ or $\varrho_{si}(t)$ during communication. Hence, only the manipulated output $\check{y}_{i}(t)=\check{x}_{i,1}(t)$ and state $\check{x}_{i,2}(t)$ are accessible to the local controller. Meanwhile, $\varrho_{ai}(t)$ is the unknown weight that differentiates the pre-designed controller $u_i(t)$ from actuator-received $\check{u}_i(t)$.

From \eqref{system model} and \eqref{2}-\eqref{4}, the dynamics of local manipulated states $\check{x}_{i}=[\check{x}_{i,1},\check{x}_{i,2}]^T$ can be further
derived as
\begin{equation}
 \label{5}
 \setlength\abovedisplayskip{6pt}
\setlength\belowdisplayskip{5pt}
 \begin{aligned}
\!\!\!\!\dot{\check{x}}_{i,1}&=\varrho_{o}\psi_{i,1}^T\vartheta_{i,1}\!+g_{i,1}\varrho_{o}\varrho_{si}^{-1}\check{x}_{i,2}+\varrho_{o}o_{i,1}+\dot{\varrho}_{o}\varrho_{o}^{-1}\check{x}_{i,1},\\
\!\!\!\!\dot{\check{x}}_{i,2}&=\varrho_{si}\psi_{i,2}^T\vartheta_{i,2}\!+g_{i,2}\varrho_{si}\varrho_{ai}u_i+\varrho_{si}o_{i,2}+\dot{\varrho}_{si}\varrho_{si}^{-1}\check{x}_{i,2}.
\end{aligned}
\end{equation}
\begin{remark}\label{remark 1}
Deception attacks formed by \eqref{2}-\eqref{4} have been studied in many studies \cite{7,8,9,10,11,12}. Such vicious behaviors usually possess the resistance to network secure detective devices \cite{17}, then bring impacts in a subtle way. From \eqref{5}, the impacts are reflected from the following two aspects:
\begin{enumerate}[(1)]
\vspace{-0.5mm}
\item Sensor weights $\varrho_{o}(t)$ or $\varrho_{si}(t)$ invite time-varying uncertain terms such as $\dot{\varrho}_{o}\varrho_{o}^{-1}\check{x}_{i,1}$  in the dynamics \eqref{5}.
\item The coefficients of $\check{x}_{i,2}$ and $u_i$ are given by $g_{i1}'=g_{i,1}\varrho_{o}\varrho_{si}^{-1}$  and $g_{i2}'=g_{i,2}\varrho_{si}\varrho_{ai}$, respectively, each with unknown signs. Therefore, it is possible for the negative-feedback signal $u_i(t)$ to be imperceptibly reversed into a positive-feedback signal $\check{u}_i(t)$ if $\operatorname{sign}(\varrho_{o}\varrho_{si}^{-1})= -1$ or $\mathrm{sign}(\varrho_{si}\varrho_{ai})=-1$, severely destabilizing the MASs. The resilient control problem under attacks is more complex than the fault-tolerant problem with efficiency-scaling faults $u_r(t)=\varpi(t) u(t), \varpi(t)>0$ (see \cite{4} as instances).
\vspace{-0.5mm}
\end{enumerate}
All the above discussions highlight the necessity of developing resilient consensus protocols to withstand attacks.
\end{remark}
\begin{remark}\label{remark sensorweights}
In \cite{9,10}, all attack weights $\varrho_o(t)$, $\varrho_{si}(t)$, and $\varrho_{ai}(t)$ are assumed to be identical across agents. In contrast, we allow $\varrho_{si}(t)$ and $\varrho_{ai}(t)$ to be non-identical, while $\varrho_o(t)$ is identical. This reflects a more realistic scenario where internal states $x_{i,2}(t)$ and inputs $u_i(t)$ are typically hard to compromise uniformly due to confidentiality, while outputs $y_i(t)$ are more exposed and easier to be uniformly modified.
\end{remark}
\subsection{Control Objective }
The \emph{control objective} is to devise distributed
adaptive controllers for MASs \eqref{System block} to guarantee the resilient consensus performance in the sense that:
\begin{enumerate}[$\bullet$]
  \item all the closed-loop signals maintain uniform boundedness even in the face of deception attacks;
  \item each output consensus error $y_i-y_j(\forall i,\!j \!\leq \! N\!)$  converges into a clearly defined neighbor-set of the origin in finite time.
\end{enumerate}

To attain the above targets, a few assumptions are stated.
\vspace{-0mm}
\begin{assumption}
$g_{i,j}(t)$ stays in a bounded interval, $i.e.$, $|g_{i,j}(t)|\! \in\! [\underline{g}, \overline{g}]$, where $\underline{g}$ and $\overline{g}$ stand for two positive constants.
\end{assumption}
\begin{assumption}
$o_{i,j}(t)$ remains in a bounded set, $i.e.$, $o_{i,j}(t) \in S_{o}\triangleq\{|o_{i,j}|\leq \bar{o}\}$, where $\bar{o}$ is an unknown constant.
\end{assumption}
\begin{assumption}
For attack weights, there exists unknown sets: $\varrho_o(t)\! \in \! S_{\!1}\!\triangleq\{|\varrho_o|\leq \bar{\varrho}_o,|\varrho_o|\geq \underline{\varrho}_o,|\dot{\varrho}_o|\leq \varrho_d \}$, $\varrho_{si}(t) \in S_{2}\triangleq\{|\varrho_{si}|\leq \bar{\varrho}_s,|\varrho_{si}|\geq \underline{\varrho}_s, |\dot{\varrho}_{si}|\leq \varrho_d \} $ and $\varrho_{ai}(t) \in S_{3}\triangleq\{|\varrho_{ai}|\leq \bar{\varrho}_a,|\varrho_{ai}|\geq \underline{\varrho}_a \} $, respectively, where $\bar{\varrho}_o, \bar{\varrho}_s, \bar{\varrho}_a, \underline{\varrho}_o, \underline{\varrho}_s, \underline{\varrho}_a, \varrho_d$ are all positive constants.
\end{assumption}
\begin{remark}\label{remark 2}
Assumption 1 ensures the controllability of system \eqref{system model} by requiring nonzero control coefficients. Assumption 2 implies the boundedness of disturbance \cite{3}. From  \cite{12}, Assumption 3 implies  that attackers deliberately manipulate attack signals to enhance the ability to evade detection.
\end{remark}

Some lemmas are exerted to prop up the control design.
\vspace{0mm}
\begin{lemma}[\!\!\cite{12}]\label{Lemma 2}
 Under Assumption 3, one can find a list of known smooth functions $\varphi_{i,1} (\check{y}_{i})$ and $\varphi_{i,2} (\check{x}_{i})$ such that $|| \psi_{i,1}(x_{i,1})|| \leq \xi_{i1}\varphi_{i,1} (\check{y}_{i})$ and $|| \psi_{i,2}(x_{i})|| \leq \xi_{i2}\varphi_{i,2} (\check{x}_{i})$ with unknown constants $\xi_{i1}$, $\xi_{i2}$.
\end{lemma}

As claimed earlier, potential attacks may lead to the change of control directions. To address this issue, we resort to the enhanced Nussbaum-type function as defined below.

\begin{definition}[\!\!\cite{15}]\label{Definition 1}
 Given $\mathcal{N}(\nu)$ as a continuously differentiable real function of variable $\nu$ ($\nu \in \mathbb{R}$), such that
\begin{equation}
\label{6}
\begin{aligned}
\!\!\!\!\!&\liminf_{w \rightarrow \infty} \frac{w\!-\!\!\int_{0}^{w}\mathcal{N}_n(\nu)d\nu}{\int_{0}^{w}\mathcal{N}_p(\nu)d\nu}\!=\!\liminf_{w \rightarrow \infty} \frac{w\!+\!\!\int_{0}^{w}\mathcal{N}_p(\nu)d\nu}{-\!\int_{0}^{w}\mathcal{N}_n(\nu)d\nu}\!=\!0,\!\!\!\!\!\!\!\!
\end{aligned}
\end{equation}
where $\mathcal{N}_p(\nu)\triangleq\max\{\mathcal{N}(\nu),0\}$, $\mathcal{N}_n(\nu)\triangleq\mathcal{N}(\nu)-\mathcal{N}_p(\nu)$, then it is termed as $\mathcal{N}$-\emph{Function} denoted as $\mathcal{N}(\nu) \in \mathcal{N}$.
\end{definition}

Then, the following lemma is given as a stability criteria.
\begin{lemma}[\!\!\cite{18}]\label{Lemma 3}
Choose a non-negative differentiable function $V(t)$ and two positive constant $J,K$, it renders that
\begin{equation}
\label{7}
\begin{aligned}
\dot{V}(t) \!\leq\! -J V(t)\!+\!K\!+\!\sum\nolimits_{i=1}^{M} \left[1\!+ \!\mu_i(t)\mathcal{N}(\digamma_i(t)) \right]\dot{\digamma_i}(t),
\end{aligned}
\end{equation}
with gains $\digamma_i(t)$ satisfying $\dot{\digamma}_i\geq 0$, coefficient $\mu_i(t)$ fulfills that
 $\mu_i(t)=\bar{\mu}_i(t)\mu_{il}+(1-\bar{\mu}_i(t))\mu_{iu}$ with constants $\mu_{il}\mu_{iu}>0$, $\bar{\mu}_i(t)\in [0,1]$. Then, $\digamma_i(t), V(t)$ are bounded, $\forall t\! \in[0,+\infty)$.
\end{lemma}

\section{Design of Distributed Adaptive Resilient Controllers}
\label{sec:DesignofDistributedAdaptiveResilientControllers}
\subsection{Finite-time Distributed Reference Systems}
Distributed reference systems are widely used to achieve consensus of nonlinear MASs \cite{9,10,11}, where output consensus is realized if all of reference states reach an agreement and each agent's output follows its own reference state. In \cite{9,10,11}, an exponential-convergent reference system are designed as: $\dot{s}_i\!=\!-\sum_{j=1}^N a_{ij}(\check{y}_i\!-\!\check{y}_j)$, where $s_i$ is the reference state of agent $i$. However, this design cannot ensure finite-time consensus of $s_i$. To achieve the control objective, we design a class of distributed finite-time reference systems:
 \begin{equation}
  \label{8}
  \begin{aligned}
  \dot{s}_i=-k|\varpi_i|^\alpha \mathrm{sign}(\varpi_i), \  \varpi_i\triangleq \sum\nolimits_{j\in \mathbf{N}_i} (s_i-s_j)
   \end{aligned}
   \end{equation}
\noindent with constants $\!k\!>\!0$, $0<\!\alpha<\!\!1$, $\varpi_i$ is the local reference error. Let $s=[s_1,\ldots,s_N]^T$, $\varpi=[\varpi_1,\ldots,\varpi_N]^T$, $\varpi=\mathcal{L}s$.

The performance of \eqref{8} is analyzed in Theorem 1.
 \vspace{0mm}
\begin{theorem}\label{theorem 1}
Consider a group of distributed reference system \eqref{8} under strongly connected topology,  $s_i(t)$ ( $i=1,\ldots,N$) attains a consensus within a finite time $T_0$. That is, there exists a constant $s_0$ such that $\forall t \in [T_0, +\infty)$, $s_i(t)=s_0$.
\end{theorem}
\begin{ProofNoIndent} Let $\varpi_\alpha=$$[|\varpi_1|^\alpha \mathrm{sign}(\varpi_1)$$,\ldots,|\varpi_N|^\alpha \mathrm{sign}(\varpi_N)]^T$. Since $\varpi=F\varpi_\alpha$, $F \triangleq \mathrm{diag}\{f_1,\ldots,f_N\}$, $ f_i\triangleq|\varpi_i|^{1-\alpha}+1-|\mathrm{sign}(\varpi_i)|$, a positive vector $l\triangleq Fh$ renders that $l^T\varpi_\alpha=0$. Then, a Lyapunov function is employed as
\begin{equation}
  \label{9}
    \begin{aligned}
  W(t)=\sum\nolimits_{i=1}^N h_i(1+\alpha)^{-1}|\varpi_i(t)|^{1+\alpha}.
    \end{aligned}
   \end{equation}
\noindent According to Lemma \ref{Lemma 1}, the derivative of $W(t)$ fulfills that
\begin{equation}
  \label{10}
    \begin{aligned}
  \!\!\!\!\dot{W}&=-k\varpi_\alpha^T\mathcal{H}\mathcal{L}\varpi_\alpha=-2^{-1}k\varpi_\alpha^T\mathcal{Q}\varpi_\alpha \leq -Q W^{\frac{2\alpha}{\alpha+1}}, \!\!\!\!\!\!\!\!
    \end{aligned}
   \end{equation}
where $Q=2^{-1}k\lambda_2(\mathcal{Q})N^{-1}[(1+\alpha)\bar{h}^{-1}]^{\frac{2\alpha}{\alpha+1}}$, $\bar{h}=\max\{h_i\}$. Integrating both sides of \eqref{10}, yields that
\vspace{0mm}
\begin{equation}
   \label{11}
\begin{aligned}
  W^{\frac{1-\alpha}{1+\alpha}}(t)\leq W^{\frac{1-\alpha}{1+\alpha}}(0)-Q(1-\alpha)(1 +\alpha)^{-1} t.
   \end{aligned}
        \end{equation}
Hence, there exists a finite time $T_0$ which is bounded by
\vspace{-1mm}
\begin{equation}
 \label{12}
 \begin{aligned}
  T_0 \leq W^{\frac{1-\alpha}{1+\alpha}}(0)(1+\alpha) [Q(1-\alpha)]^{-1}
   \end{aligned}
      \end{equation}
such that $W(t)=0$ for $t>T_0$. It implies that $\omega_i(t)=0$ and $s_i(t)=s_j(t) (1\leq i,j \leq N)$ for $t>T_0$. From \eqref{8}, it also shows that $\dot{s}_i\!=\!0$, thus $s_i(t)\!=\!s_0\!\triangleq \!s_i(T_0)$, $\forall t \in [T_0, +\infty)$.
\end{ProofNoIndent}

Different from \cite{9,10,11}, $\check{y}_i$ is not involved in \eqref{8}. Otherwise, if $\dot{s}_i\!=-k| \sum\nolimits_{j\in \mathbf{N}_i} (\check{y}_i-\check{y}_j)|^\alpha \mathrm{sign}( \sum\nolimits_{j\in \mathbf{N}_i} (\check{y}_i-\check{y}_j))=-k|\varpi_i+ \sum\nolimits_{j\in \mathbf{N}_i} (e_i-e_j)|^\alpha \mathrm{sign}( \varpi_i+ \sum\nolimits_{j\in \mathbf{N}_i} (e_i-e_j))$, $e_i=\check{y}_i-s_i$ hinders finite-time convergence of $W(t)$ in \eqref{11}. Instead, as in \cite{3}, each agent $i$ sends $s_i$ to its neighbors and uses $\varpi_i$ in \eqref{8}, ensuring that Theorem 1 holds. Then, control objectives can be achieved as long as the subsequent design ensures finite-time bounded tracking of $\check{y}_i$ to $s_i$.
\begin{remark}\label{remark 3}
The consensus value $s_0$, though related to $s(0)$ and the topology, is not easy to compute explicitly, similar to the case in finite-time research \cite{4}. It is known that $s_0$ tends to $h^T\!\!s(0)\! \! $ ($i.e.$,\! the case in \cite{9,10,11}) as $\alpha\!$ approaches 1.
\end{remark}

\vspace{-0.4mm}
\subsection{Distributed Resilient Controllers}
Based on \eqref{8}, we further raise a backstepping-based adaptive resilient controller. We adopt an error transformation as
\begin{equation}
 \label{13}
   z_{i1}=L_ie_i=L_i(\check{y}_i-s_i), \ \   z_{i2}=\check{x}_{i,2}-\upsilon_{i1},
      \end{equation}
\noindent where $\upsilon_{i1}$ is the virtual controller to be established, $e_i$ is the inner tracking error between $\check{y}_i$ and $s_i$, $L_i$ is an
adaptive gain to regulate $e_i$, which is generated by
\begin{equation}
  \label{14}
   \begin{aligned}
   \dot{L}_i=\gamma_i \max\{e_i^2-\varsigma_i\epsilon_i^2, 0\},\  L_i(0)\geq 1
   \end{aligned}
         \end{equation}
with design parameters $\!\gamma_i>0$, $\epsilon_i>0$, $0<\!\varsigma_i\!<1$. Clearly, $L_i(t)\!\geq \!\ L_i(0)\geq 1$. The controller $u_i$ is designed by two steps.

\noindent \emph{Step i,1:} Define a Lyapunov function candidate as $V_{i1}=0.5z_{i1}^2$. From \eqref{5}, \eqref{13} and \eqref{14}, it is derived that
\begin{equation}
 \label{15}
\dot{V}_{i1}\!=\!L_iz_{i1}[\dot{L}_iL_i^{-2}z_{i1}\!+\!g_{i1}'(z_{i2}\!+\!\upsilon_{i1})\!+\!\Gamma_{i1}],
         \end{equation}
where $\Gamma_{i1}=\varrho_{o}(\psi_{i,1}^T\vartheta_{i,1}\!+\!o_{i,1})+\dot{\varrho}_{o}\varrho_{o}^{-1}\check{x}_{i,1}-\dot{s}_i$. By using Lemma \ref{Lemma 2} and Assumptions 1 to 3, we deduce that
\begin{equation}
   \label{16}
   \begin{aligned}
|L_iz_{i1}\psi_{i,1}^T\varrho_{o}\vartheta_{i,1}| \leq 4^{-1} (L_iz_{i1})^2\varphi_{i,1}^2+ \xi_{i1}^2\vartheta_{i,1}^2\bar{\varrho}_o^2,
\end{aligned}
    \end{equation}
\begin{equation}
 \label{17}
 \begin{aligned}
\!\!|&L_iz_{i1}(g_{i1}'z_{i2}+\varrho_{o}o_{i,1}+\dot{\varrho}_{o}\varrho_{o}^{-1}\check{x}_{i,1})|\leq 4^{-1}(L_iz_{i1}z_{i2})^2  \\
&+4^{-1}(L_iz_{i1})^2(1+\check{x}_{i,1}^2)+\overline{g}^2\bar{\varrho}_o^2\underline{\varrho}_s^{-2}+ \bar{\varrho}_o^2\bar{o}^2+ \varrho_d^2\underline{\varrho}_o^{-2}.
\end{aligned}
\end{equation}
From Theorem 1, there is a constant $\bar{s}_d$,$s.t.$,$|\dot{s}_i|\leq \bar{s}_d$, then
\begin{equation}
 \label{18}
 \begin{aligned}
|L_iz_{i1}\dot{s}_i|\leq 4^{-1}(L_iz_{i1})^2+\bar{s}_d^2.
   \end{aligned}
   \end{equation}
Taking \eqref{16}$-$\eqref{18} into \eqref{15}, yields that
\begin{equation}
 \label{19}
  \begin{aligned}
\dot{V}_{i1}\leq& L_iz_{i1}\left[4^{-1}L_iz_{i1}(\varphi_{i,1}^2+2+\check{x}_{i,1}^2)+g_{i1}'\upsilon_{i1} \right. \\
&\left.\!\!+\dot{L}_iL_i^{-2}z_{i1}\right]+4^{-1}(L_iz_{i1}z_{i2})^2+\Lambda_{i1},
   \end{aligned}
      \end{equation}
where $\Lambda_{i1}=\xi_{i1}^2\vartheta_{i,1}^2\bar{\varrho}_o^2+\overline{g}^2\bar{\varrho}_o^2\underline{\varrho}_s^{-2}+ \bar{\varrho}_o^2\bar{o}^2+ \varrho_d^2\underline{\varrho}_o^{-2}+\bar{s}_d^2$ is an unknown constant.
Based on Definition \ref{Definition 1}, $\upsilon_{i1}$ is drawn as
\begin{equation}
\setlength\abovedisplayskip{6pt}
\label{20}
\upsilon_{i1}\!\!=\!\mathcal{N}(\digamma_{i1}\!)[c_{i1}e_i + \tfrac{L_iz_{i1}}{4}(\varphi_{i,1}^2\! +2 + \check{x}_{i,1}^2)+\gamma_i (e_i^2+\epsilon_i^2)\tfrac{z_{i1}}{L_i^{2}}]\!
      \end{equation}
with Nussbaum gain $\digamma_{i1}$ determined by
\begin{equation}
\label{21}
\setlength\abovedisplayskip{6pt}
\!\!\!\dot{\digamma}_{i1}\!=\!L_iz_{i1}[c_{i1}e_i\!+\!\tfrac{L_iz_{i1}}{4}(\varphi_{i,1}^2+2+\check{x}_{i,1}^2)\!+\!\gamma_i (e_i^2+\epsilon_i^2)\tfrac{z_{i1}}{L_i^{2}}],\!\!\!
      \end{equation}
where $c_{i1}>0$ is a design parameter. Then, there is
\begin{equation}
\label{22}
 \begin{aligned}
\!\!\!\!\!\!\!\dot{V}_{i1}\!\leq \! \!-c_{i1}z_{i1}^2\!+\![g_{i1}'\mathcal{N}(\digamma_{i1})\!+\!1]\dot{\digamma}_{i1}\!\!+\! \Lambda_{i1}\!+\!4^{-1}(L_iz_{i1}z_{i2})^2\!. \!\!\!\!\!
   \end{aligned}
      \end{equation}
\noindent \emph{Step i,2:}  From \eqref{13} and \eqref{20}, $\upsilon_{i1}$ is inferred as a continuously differentiable function decided by $\check{x}_{i,1}$, $s_i$, $L_i$ and $\digamma_{i1}$. Defining a Lyapunov function as $V_{i2}=0.5z_{i2}^2$, we have
\begin{align}
\label{23}
\!\!\!&\!\!\!\dot{V}_{i2}\!=\!z_{i2}[\varrho_{si}(\psi_{i,2}^T\vartheta_{i,2}\!+\!o_{i,2}\!)\!+\!g_{i2}'u_i\!+\!\!\dot{\varrho}_{si}x_{i,2} \!-\!\tfrac{\partial \upsilon_{i1}}{\partial s_i} \dot{s}_i\!-\!\tfrac{\partial \upsilon_{i1}}{\partial L_i}\dot{L}_i  \notag \\
&-\tfrac{\partial \upsilon_{i1}}{\partial \check{x}_{i,1}}\!(\varrho_{o}(\psi_{i,1}^T\vartheta_{i,1}\!\!+\!o_{i,1}\!)\!+\!g_{i1}'\check{x}_{i,2}\!+\!\dot{\varrho}_{o}x_{i,1}\!) \!-\!\tfrac{\partial \upsilon_{i1}}{\partial \digamma_{i1}}\dot{\digamma}_{i1}\!].\!
\end{align}
Taking an analyzing process similar to \emph{Step i,1}, we have
\begin{equation}
\label{24}
\setlength\abovedisplayskip{5pt}
\setlength\belowdisplayskip{5pt}
\begin{aligned}
\dot{V}_{i2}\leq z_{i2}(z_{i2}\Phi_{i2}+g_{i2}'u_i)+\Lambda_{i2},
\end{aligned}
 \end{equation}

 \begin{table}[!t]
	\centering  
	\caption{Resilient control design of agent $i$}  
\vspace{-1mm}
	\label{table1}  
\begin{tabular}{l}
		\hline  
	  \textbf{$\bullet$ Reference System:} $\dot{s}_i=-k|\varpi_i|^\alpha \mathrm{sign}(\varpi_i).$\\ 
      \textbf{$\bullet$ Error Variables:} $z_{i1}=L_i(\check{y}_i-s_i), z_{i2}=\check{x}_{i,2}-\upsilon_{i1}$.\\
  \textbf{$\bullet$ Control Input:} $ u_{i}\!=\!\mathcal{N}\!(\digamma_{\!\!i2})[c_{i2}z_{i2}\!\!+\!\!z_{i2}\Phi_{i2}\!\!+\!4^{\!-\!1}\!L_iz_{i1}^2z_{i2}].$\\
   \textbf{$\bullet$ Update Laws:} $\dot{L}_i\!=\!\gamma_i \max\{e_i^2-\varsigma_i\epsilon_i^2, 0\}$, $L_i(0)\geq 1$,\\
    \ \ $\dot{\digamma}_{\!i1}\!\!=\!L_iz_{i1}[c_{i1}e_i\!+\!\!\frac{1}{4}L_iz_{i1}(\varphi_{i,1}^2\!\!+\!\!2\!+\!\!\check{x}_{i,1}^2)\!+\!\gamma_i (e_i^2\!+\!\epsilon_i^2)\frac{z_{i1}}{L_i^{2}}]$, \\
      \ \ $\dot{\digamma}_{\!\!i2}\!=\!z_{i2}[c_{i2}z_{i2}\!+\!z_{i2}\Phi_{i2}+4^{-1}L_iz_{i1}^2z_{i2}]$.  \\
        \hline
	\end{tabular}
\vspace{-2mm}
\end{table}
\begin{figure}[t]
\setlength{\belowcaptionskip}{-0.4cm}
\centering
\includegraphics[width=3.8cm,height=0.8cm]{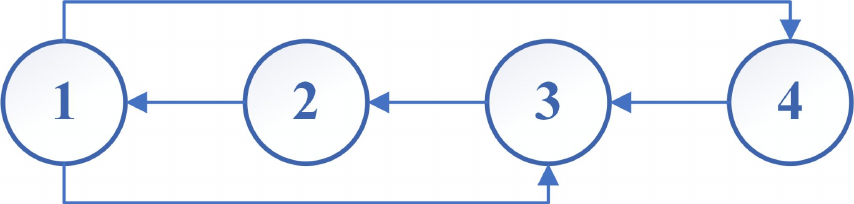} 
\caption{Communication topology among 4 agents.}
\label{fig2}
\end{figure}

\noindent where $\Phi_{i2}(\check{x}_{i},s_i,L_i,\digamma_{i1})$ denotes a continuous function and $\Lambda_{i2}$ is an unknown constant, which are defined below.
\begin{equation}
\label{25}
\setlength\belowdisplayskip{5pt}
\begin{aligned}
\Phi_{i2}=&\tfrac{(||\varphi_{i,2}||^2+1+\check{x}_{i,2}^2) }{4}+(\tfrac{\partial \upsilon_{i1}}{\partial s_i})^2+(\tfrac{\partial \upsilon_{i1}}{\partial \digamma_{i1}}\dot{\digamma}_{i1})^2\\
&+(\tfrac{\partial \upsilon_{i1}}{\partial \check{x}_{i,1}})^2\tfrac{(||\varphi_{i,1}||^2+1+\check{x}_{i,1}^2+\check{x}_{i,2}^2)}{4}+(\tfrac{\partial \upsilon_{i1}}{\partial L_i} \dot{L}_i)^2 ,
\end{aligned}
\end{equation}
\begin{equation}
\label{26}
\begin{aligned}
\Lambda_{i2}=\xi_{i2}^2\vartheta_{i,2}^2\bar{\varrho}_s^2+\bar{\varrho}_s^2\bar{o}^2+\varrho_d^2\underline{\varrho}_s^{-2}+\Lambda_{i1}+0.5.
\end{aligned}
\end{equation}
According to Definition 1, $u_i$ is drawn as
\begin{align}
u_{i}&=\mathcal{N}(\digamma_{i2})[c_{i2}z_{i2}\!+\!z_{i2}\Phi_{i2}+4^{-1}L_iz_{i1}^2z_{i2}], \label{27} \\
\dot{\digamma}_{i2}&=z_{i2}[c_{i2}z_{i2}\!+\!z_{i2}\Phi_{i2}+4^{-1}L_iz_{i1}^2z_{i2}],\label{28}
\end{align}
where $c_{i2}>0$ is a design paramter. Then, there is
\begin{equation}
\setlength\abovedisplayskip{5pt}
 \label{29}
  \begin{aligned}
\!\!\!\!\dot{V}_{i2}\leq  \!-c_{i2}z_{i2}^2\!+\![g_{i2}'\mathcal{N}(\digamma_{i2})\!+\!1]\dot{\digamma}_{i2}\!+\! \Lambda_{i2}\!-\tfrac{(L_iz_{i1}z_{i2})^2}{4}.\!\!\!\!
   \end{aligned}
   \end{equation}

\begin{figure*}[!t]  
\centering
\setlength{\abovecaptionskip}{-0.1cm}
\setlength{\belowcaptionskip}{-0.4cm}
\begin{minipage}[t]{0.325\textwidth}
    \centering
    \includegraphics[width=\linewidth,height=4.2cm]{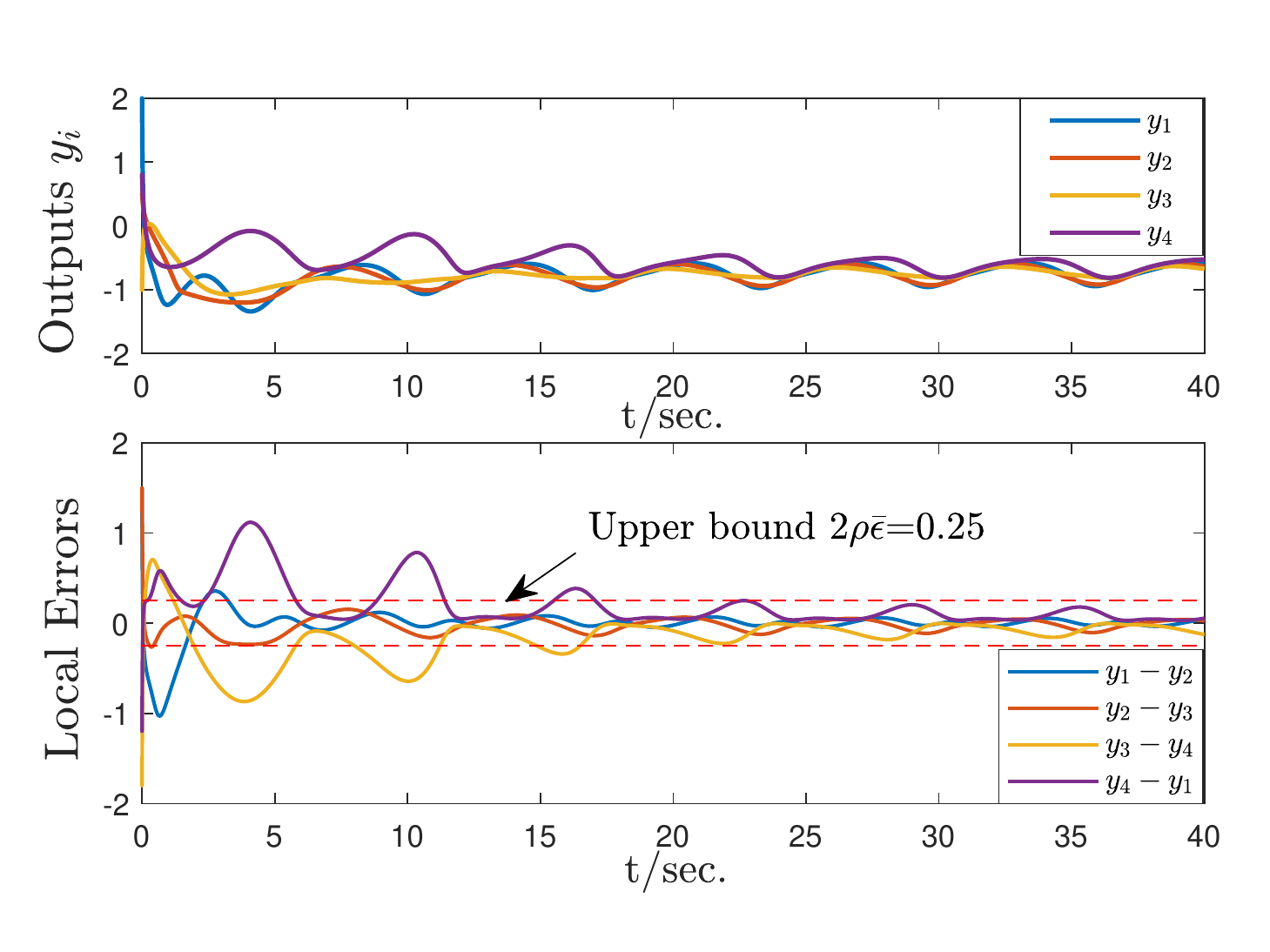}
    \caption{Outputs $y_i$ and local errors.}
    \label{fig3}
\end{minipage}
\hfill
\begin{minipage}[t]{0.325\textwidth}
    \centering
    \includegraphics[width=\linewidth,height=4.2cm]{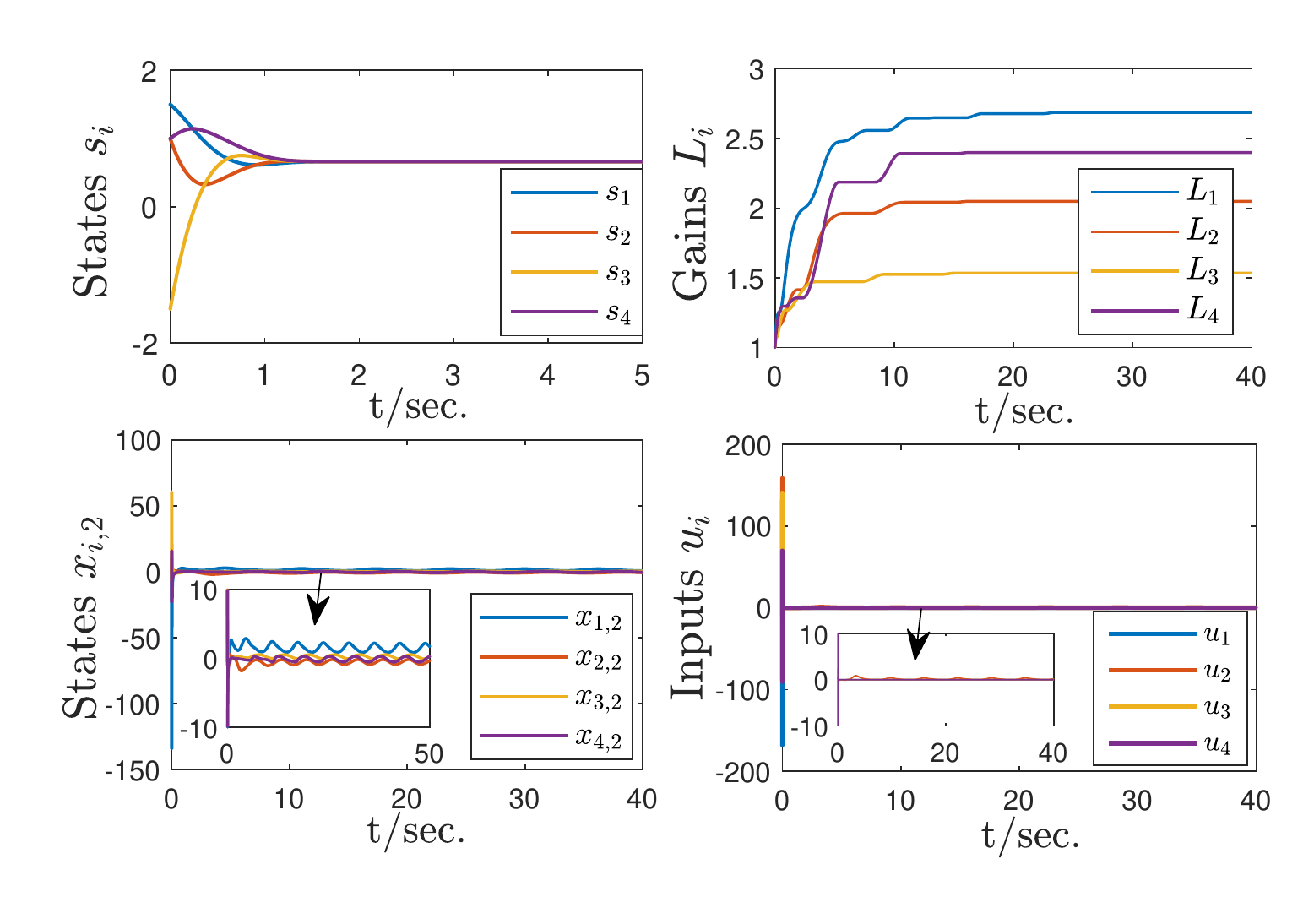}
    \caption{States $s_i$, $x_{i,2}$, $L_i$ and inputs $u_i$.}
    \label{fig5}
\end{minipage}
\hfill
\begin{minipage}[t]{0.325\textwidth}
    \centering
    \includegraphics[width=\linewidth,height=4.15cm]{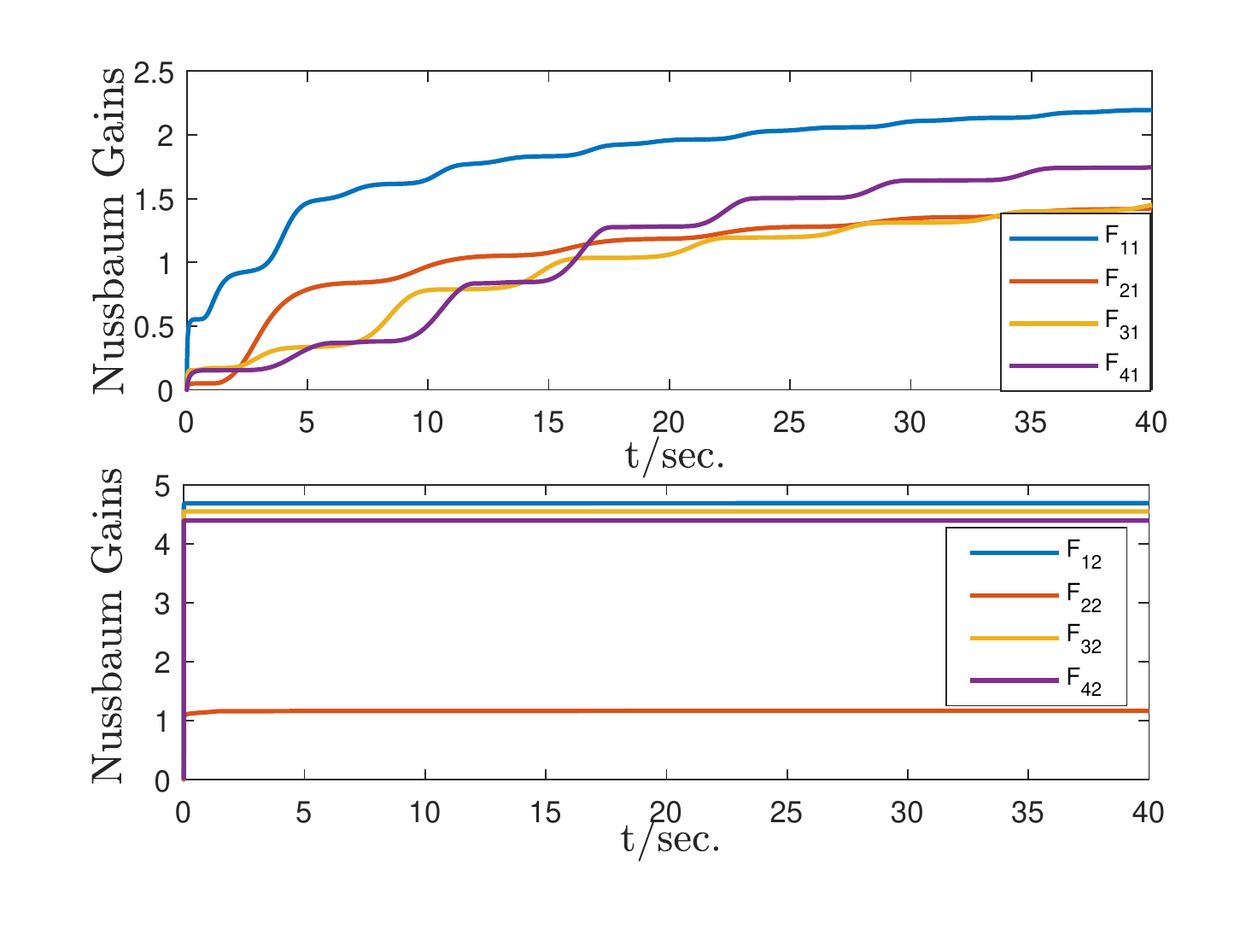}
        \caption{Nussbaum gains $\digamma_{ij}$.}
    \label{fig6}
\end{minipage}
\end{figure*}

\vspace{-3mm}
\subsection{ Stability Analysis}
Now, we summarize the control design in TABLE \ref{table1}. The system performance is specified in the following Theorem.
\begin{theorem}\label{theorem 2}
Consider a group of second-order nonlinear MASs \eqref{system model} under \emph {strongly connected} topology and deception attacks \eqref{2}, \eqref{3}, \eqref{4}. Suppose that \emph{Assumptions 1-3 hold}. The distributed reference systems \eqref{8} and controllers \eqref{27} with update laws \eqref{14}, \eqref{21}, \eqref{28} shown in TABLE \ref{table1} ensure the following resilient consensus performance for $t\in[0,+\infty)$:
\begin{enumerate}[(i)]\vspace{-0.5mm}
\item All of closed-loop signals keep uniformly bounded;
\item Within a finite time $T_s$, output consensus errors $y_i-y_j$ converge to a residual set $\Omega\!\triangleq \!\{|y_i\!-\!y_j |\!\!\leq\! \!2\rho\bar{\epsilon}, \rho \! \triangleq\underline{\varrho}_o^{-1}\!, \!\bar{\epsilon}\! \triangleq \! \max\{\epsilon_i\}\}$ whose size can be reduced by tuning $\bar{\epsilon}$.
\end{enumerate}
\end{theorem}

\begin{ProofNoIndent}
Define a total Lyapunov function as $V_i=V_{i1}+V_{i2}$. From \eqref{22} and \eqref{29}, it follows that for any $t\in[0,+\infty)$,
\begin{equation}
\setlength\abovedisplayskip{6pt}
\setlength\belowdisplayskip{6pt}
\begin{aligned}
\!\!\!\!\!\dot{V}_i(t)\!\leq\! -J_iV_i(t)\!+\!\Lambda_i\!+\!\!\sum\nolimits_{r=1}^2 \! \left(g_{ir}'\mathcal{N}(\digamma_{ir}(t))\!+\!1\right)\!\dot{\digamma}_{ir}(t), \!\!\!\!\!\label{30}
\end{aligned}
\end{equation}
\noindent where $J_i=2\min\{c_{i1},c_{i2}\}$, $\Lambda_i=\Lambda_{i1}+\Lambda_{i2}$. By using Lemma \ref{Lemma 3}, from \eqref{30}, the boundedness of $z_{ir}(t)$ and $\digamma_{ir}(t)(r=1,2)$ are ensured (note as $|z_{ir}(t)|\leq M_i$).
From \eqref{13}, $|e_i|\leq M_iL_i^{-1}$, thus $e_i(t)$ and $\dot{L}_i$ are bounded, $L_i(t)$ will not escape to infinity in finite time. Since $L_i$ is non-decreasing, assume that $\lim_{t\rightarrow+\infty} L_i(t)\!=\!+\infty$, then $\exists t_i, s.t.\ L_i(t)>M_i(0.5\varsigma_i\epsilon_i^2)^{-0.5}, \forall t>t_i$. It implies $|e_i^2(t)|-\varsigma_i\epsilon_i^2<0, \forall t>t_i$, which means from \eqref{14} that $L_i(t)=L_i(t_i), \forall t>t_i$. This contradicts the above assumption. Hence, $L_i(t)$ is bounded. Due to the boundedness of $s_i$, $\check{x}_{i,1}$ is bounded. From \eqref{13} and \eqref{21}, $\upsilon _{i1},\check{x}_{i,2}$ are bounded. Assumption 3 ensures that $x_i$ are bounded. Finally, from \eqref{27}, $u_i$ is bounded. Thus, (i) holds.

Besides, $\dot{L}_i$ is uniformly continuous. From Barbalat$'$s Lemma, $\lim_{t\rightarrow+\infty} \dot{L}_i(t)\!\!\!=\!\!\!0$. Hence, there exists a finite time $T_i$, $|e_i(t)|\!\!<\!\!\epsilon_i, \forall t\!\!>\!\!T_i$. From Theorem
\ref{theorem 1}, $|y_i(t)\!-\!y_j(t)|\leq \varrho_{o}^{-1}\! |e_i(t)-e_j(t)|\!\! \leq \!\!2\rho\bar{\epsilon}\ $ for $t\!\!>\!\!T_s\!\triangleq \!\max_{1\leq i\leq N}\{T_0, \!T_i\}$. Clearly, for a certain attack, $\Omega$ can be reduced by decreasing $\bar{\epsilon}$.
\end{ProofNoIndent}
\begin{remark} \label{remark 4}
In \cite{10}, distributed consensus controllers are built by defining $z_{i1}=\check{y}_i-s_i$. They can merely ensure the boundedness of consensus errors, but the expression of these upper bounds is elusive, and so are the relationships between control parameters and the upper bound, because the bound of Nussbaum terms $(g_{ir}'\mathcal{N}(\digamma_{ir})+1)\dot{\digamma}_{ir}$ as in \eqref{30} is unknown. It results in unknown upper bounds of $V_i(t)$. To overcome this drawback, an adaptive gain $L_i$ is incorporated in $z_{i1}$. It forces $e_i$ to converge into a set $\{|e_i|\leq\epsilon_i\}$. Hence, the residual set $\Omega$ is sketched clearly. The parameter $\rho$ stands for the sensor attack intensity. By reducing $\bar{\epsilon}$, even if attackers increase $\rho$, $2\rho\bar{\epsilon}$ may be unchanged, $i.e.$, the system's resilience to attacks can be maintained. For a certain $\rho$, the consensus degree of $y_i$ can be improved by reducing $\bar{\epsilon}$. Then, $T_s$ can be reduced by increasing $k$ or $\gamma_i$. Besides, Theorem 2 shows that $|y_i(t)-\varrho_{o}^{-1}(t)s_0|\leq \rho\bar{\epsilon}, \forall t\!>\!T_s$. Remark \ref{remark 3} shows that the consensus value $\varrho_{o}^{-1}(t)s_0$ of $y_i(t)$ is time-varying and related to attack weight $\varrho_{o}(t)$, the initial value $s(0)$ and the topology.
\end{remark}
\begin{remark} \label{remark 5}
In \cite{9} and \cite{11}, sign-approximation functions $s_g(e_{i})\!=\!e_{i}(e_{i}^2\!+\!\exp(-bt))^{-0.5}(b\!>\!0)$ are engaged in $\upsilon_{i1}$ to manage the time-varying uncertainties. Then, $\Lambda_i$ in \eqref{30} is replaced by $\exp(-bt)$ so that $y_i$ achieves asymptotic consensus. However, as a cost, as $t$ tends to infinity, the fact that $e_{i}$ tends to 0 renders $\frac{\partial s_g(e_{i})}{\partial e_i}$ becomes unbounded. Thus, $\frac{\partial \upsilon_{i1}}{\partial x_{i,1}}$ in their second virtual controller may be unbounded, so $x_{i,2}(t)$ cannot be strictly ensured to be bounded over the entire time interval $[0,+\infty)$. In contrast, the designed controllers strictly ensure that all the closed-loop signals are bounded for $t\in[0,+\infty)$.
\end{remark}
\begin{remark} \label{remark 6}
Some inspirations for future works can be sketched. Firstly, the settling time $T_s$ is not precisely formulated in this scheme, which needs more attentions. Then, it is promising to adopt barrier Lyapunov
functions \cite{revier3} to mitigate the output performance
degradation. Besides, event-triggered mechanism \cite{revier3} can be used to reduce communication burden.
\end{remark} \label{remark 6}
\subsection{A Further Discussion of Nussbaum Functions}
As discussed in \cite{14,15}, under the possibility of non-identical unknownness of multiple control directions, universal Nussbaum functions ($e.g.,\ N(\nu)=\sin(\nu) \nu^2$, $\nu \in \mathbb{R}$) fail to stabilize a system. $\mathcal{N}$-\emph{Function} \eqref{6} has been proven effective in addressing this issue in \cite{11,15}. Its mechanism is to use $\digamma_{ij}$ to extract system information and adaptively identify unknown control directions. As $\digamma_{ij}$ grows, controllers alternate feedback signs until correct directions are identified, after which $u_i$ remains in negative feedback and stabilize closed-loop systems. Hence, we use \eqref{6} to counteract deception attacks.

However, so far, only a class of functions $\!\mathcal{N}\!(\nu)\!=e^{a\nu^2}\!\sin(b\nu\!+\!c\pi/2) $
\!\!$(a\!>\!0, b,c\! \in \mathbb{R})$ has been proved as $\mathcal{N}$\emph{-Function} in \cite{15}.
To broaden the applicability, we certify two generallized forms of $\mathcal{N}$-function in Theorem 3.
 \begin{theorem}\label{theorem 3}
 Consider a real function formed as
\begin{equation}
\setlength\abovedisplayskip{5pt}
\setlength\belowdisplayskip{5pt}
\label{31}
\begin{aligned}
\!\!(1)\ \mathcal{N}(\nu)=e^{f(\nu)}\sin(\omega\nu); \ (2)\ \mathcal{N}(\nu)=e^{f(\nu)}\cos(\omega\nu),
\end{aligned}
\end{equation}
where $\omega>0$ and $\nu \in \mathbb{R}$, $f(\nu)$ is a continuously differentiable even function, $f(\nu)$ and $f'(\nu)$ keep strictly increasing for $\!\nu\!>\!0$, $\lim_{\nu \rightarrow +\infty}f(\nu)\!=\!\lim_{\nu \rightarrow +\infty}f'(\nu)\!=\!+\infty$, \!then it shows that:
\begin{enumerate}[(i)]
\item $\mathcal{N}(\nu)$ is a $\mathcal{N}$-function;
\item  The proposed resilient control objectives can be achieved by adopting $\mathcal{N}(\nu)$ in controllers \eqref{27}.
\end{enumerate}
 \end{theorem}
\ \ \ \emph{Proof:} See the Appendix I.

Theorem 3 gives a broader framework for selecting $\mathcal{N}$-functions. When the form in \cite{15} is used in controllers \eqref{20} and \eqref{27}, the non-decreasing of $\digamma_{ij}(t)$  may cause rapid growth of $e^{a\digamma_{ij}^2}$, leading to output oscillations or spikes and degrading dynamic performance (see simulations in \cite{9,11}). In contrast,  the form \eqref{31} allows for slower-growing functions such as $\mathcal{N}(\nu)\! =\! e^{a(\nu^2 + b)^c}\!\! \sin(l\nu)$ ($a \!\!> \!\!0$, $b \!\geq \!0$, $0.5\! < \!c\!\! <\!\! 1$, $\!l \! \in\! \mathbb{R}$), which mitigates such effects while ensuring output consensus.

\section{Simulation Example}
\label{sec:Simulation}
Consider a group of MASs with communication topology depicted in Fig. \ref{fig2}. The dynamics are described by \eqref{system model} with $\vartheta_{1,1}\!\!=\!-\vartheta_{1,2}\!=\!2$, $\vartheta_{2,1}\!\!=\!\!-\vartheta_{2,2}\!=\!-1$, $\vartheta_{3,1}\!\!=\!-\vartheta_{3,2}\!=\!0.5$, $\vartheta_{4,1}\!=\!\vartheta_{4,2}\!=\!-0.2$, $\psi_{1,1}\!=\!x_{1,1}$, $\psi_{1,2}\!=\!x_{1,1}\!x_{1,2}$, $\psi_{2,1}\!=\!x_{2,1}\!\sin(x_{2,1})$, $\!\psi_{2,2}\!\!=\!x_{2,2}$, $\psi_{3,1}\!\!=\!x_{3,1}\!\cos(x_{3,1})$, $\psi_{3,2}\!\!=\!x_{3,2}\tanh(x_{3,2})$, $\psi_{4,1}\!\!=\!x_{4,1}$, $\psi_{4,2}\!\!=\!\!x_{4,1}\!x_{4,2}$, $g_{1,1}\!\!=\!\!g_{1,2}\!\!=\!\!g_{3,1}\!\!=\!\!-g_{3,2}\!\!=\!\!1$, $g_{2,1}\!\!=\!\!\!-1\!-\!0.2\cos(t)$, $g_{2,2}\!\!=\!2\!-0.5\!\cos(t)$, $g_{4,1}\!=\!g_{4,2}\!=\!2\!-\!1.2\cos(t)$, $o_{1,1}\!\!=\!o_{1,2}\!\!=\!-o_{3,1}=\!-o_{3,2}=\!0.4\!\sin(t)$, $o_{2,1}=o_{2,2}=-o_{4,1}=-o_{4,2}\!=0.4\cos(t)$. Attacks are $\varrho_{o}\!=\!-1\!-0.2\sin(t)$, $\varrho_{s1}=\!1+\!0.5\sin(2t)$, $\varrho_{s2}\!=\!-1-0.3\sin(2t)$, $\varrho_{s3}\!=2-1.2\cos(2t)$, $\varrho_{s4}\!=\!2-1.5\cos(2t)$, $\varrho_{a1}\!=1\!+0.4\cos(4t)$, $\varrho_{a2}\!=1\!-0.3\sin(2t)$, $\varrho_{a3}\!=\!-2+1.4\cos(2t)$, $\varrho_{a4}\!=\!2-1.4\cos(4t)$.

Firstly, we use function $\mathcal{N}(\nu)\!=\!e^{0.9(\nu^2+0.001)^{0.7}}\!\cos(0.4\nu)$. Other parameters and initial values are $k\!\!=\!\!2$, $\alpha\!\!=\!\!0.8$, $\!c_{i1}\!\!=\!\!c_{i2}\!\!=\!\!0.2$, $\epsilon_i\!\!=\!0.1$, $\!\varsigma_i\!\!=\!0.9$, $\gamma_i\!\!=\!\!1.5$, $[x_{1,1}(0),x_{1,2}(0),$ $ x_{2,1}(0), x_{2,2}(0),$$ x_{3,1}(0), x_{3,2}(0), x_{4,1}(0),$$ x_{4,2}(0)]=[2,-1,$ $0.5,-0.5,-1,1,0.8,-1]$, $[s_1(0),s_2(0),s_3(0),s_4(0)]=[1.5,$ $1,-1.5,1]$, $L_i(0)\!=\!1$, $\digamma_{ij}(0)\!=0$. Fig. \ref{fig3}$-$\ref{fig6} show that all the closed-loop signals are bounded and consensus errors enter the set $\Omega$ within a finite time $T\!\!<\!\!25$s. Then, we choose different $\bar{\epsilon}$ and show the sum of error squares (\emph{ES Sum}) $E\!\!=\!\!\sum_{i,j\leq N} || y_i-y_j||^2$ in Fig. \ref{fig7}(a). $E$ gets smaller as $\bar{\epsilon}$ increases. From Fig. \ref{fig7}(b), $L_i$ drives $y_1\!-y_4$ into the $\bar{\epsilon}$-dependent set $\Omega$, which is hard to attained by method \cite{10} ($L_i\!\equiv1$). From  Fig. \ref{fig7}(c), compared to the performance under $\mathcal{N}(\nu)\!\!=\!\!e^{\nu^2}\!\!\!\cos(0.4\nu)$ in \cite{15}, the proposed $\mathcal{N}$\emph{-Function} can reduce spikes.

\section{Conclusion}
\label{sec:Conclusion}
In this paper, a novel distributed resilient consensus control
protocol is developed for nonlinear MASs to counter local deception attacks. By engaging enhanced Nussbaum functions and verifying a generalized form, the challenges posed by multiple unknown control directions resulting from attacks are effectively addressed, and the following output performance degradation can be mitigated. Using finite-time distributed reference systems and adaptive gains, this method ensures bounded output consensus while maintaining the overall closed-loop stability.
Simulation results show its effectiveness.
\begin{figure}[t]
\centering
\setlength{\abovecaptionskip}{-0.8cm}
\setlength{\belowcaptionskip}{-0.4cm}
\includegraphics[width=\columnwidth,height=5.8cm]{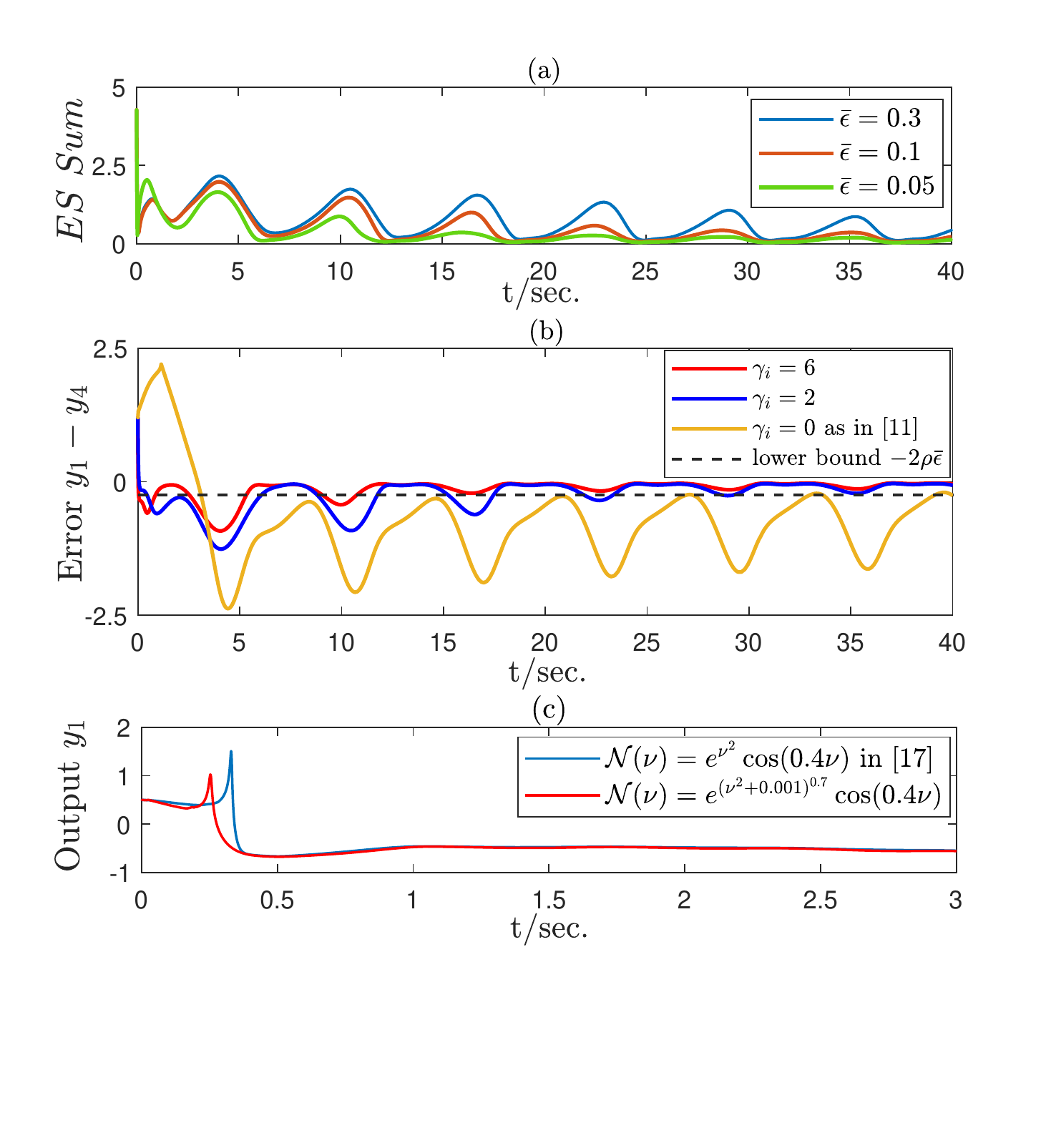} 
\caption{(a) E under different $\bar{\epsilon}$; (b) Error $y_1-y_4$  under different $\gamma_i$; (c) Output $y_1$ under different Nussbaum functions.}
\label{fig7}
\end{figure}

\appendices
\renewcommand{\theequation}{\Alph{section}.\arabic{equation}}
\section{}

\setcounter{equation}{0}
\begin{ProofNoIndent}
(1). \emph{Case 1($w\!>\!0$):} Denote $T=2\pi\omega^{-1}, w_{i}=0.5iT, i\in \mathbb{N}^+$. Define two sequences $\{Z_i^{p}\}$ and $\{Z_i^{n}\}$ as
 \begin{align}
\!\!\!\!Z_i^{p}&\!= \! \int_{w_{2i-2}}^{w_{2i-1}}\!\!\mathcal{N}(\nu)d\nu=\int_{0}^{0.5T}\!\!e^{f(r+w_{2i-2})}\sin(\omega r)dr,\!\! \!\label{A1} \\
\!\!\!\!Z_i^{n}\!&= \!\!  \int_{w_{2i-1}}^{w_{2i}}\!\!\mathcal{N}(\nu)d\nu=\!\!-\!\!\int_{0}^{0.5T}\!\!e^{f(r+w_{2i-1})}\sin(\omega r)dr.\!\!\!\!\! \label{A2}
\end{align}
 It is clear that $Z_i^{p}>0$, $Z_i^{n}<0$. Since $f(\nu)$ is strictly increasing, there is $Z_{i+1}^{p}>Z_{i}^{p}$ and $|Z_{i+1}^{n}|>|Z_{i}^{n}|$.

Firstly, we consider the sequence $\{w_{2i-1}\}$. It follows that
 \begin{equation}
 \begin{aligned}
  \label{A3}
\!\!\!\! \int_{0}^{w_{2i-1}}\!\!\!\!\!\mathcal{N}_p(\nu)d\nu\!\!=\!\!\!\sum\nolimits_{k=1}^{i}\!\!\!\! Z_{k}^{p}\triangleq \!S_i^p, \! \int_{0}^{w_{2i-1}}\!\!\mathcal{N}_n(\nu)d\nu\triangleq S_i^n.
\end{aligned}
\end{equation}
\noindent It is clear that $S_i^p\!\!>\!\!iZ_{1}^{p}$, so $\lim_{i \rightarrow +\infty }\!S_i^p\!\!=\!+\infty$. Based on the Mean-value Theorem and the strict increasing of $f'(\nu)$, it shows $f(r+w_{2i-2})\!-\!f(r\!+\!w_{2i-3})\!>\!f'(w_{2i-3})\frac{T}{2}, \forall r \in [0,\frac{T}{2}]$, leading to
\begin{equation*}
\begin{aligned}
\frac{-Z_{i-1}^n}{Z_i^p}&\!=\frac{\int_{0}^{0.5T}\!\!e^{f(r+w_{2i-3})}\sin(\omega r)dr}{\int_{0}^{0.5T}\!\!\!\!e^{f(r+w_{2i-3})}e^{f(r+w_{2i-2})-f(r+w_{2i-3})}\sin(\omega r)dr} \\
&\leq \!e^{-0.5T f'(w_{2i-3})}
\end{aligned}
\end{equation*}
Hence, from Stolz Theorem \cite{20}, we can deduce that
\begin{equation}
 \begin{aligned}
 \label{A4}
\!\!\!\!\!\!\!\lim_{i \rightarrow\! +\infty }\!\!\frac{-S_i^{n}}{S_i^{p}}\!=\!\lim_{i \rightarrow +\infty }\frac{-(S_i^{n}-S_{i-1}^{n})}{S_i^{p}-S_{i-1}^{p}}\!=\!\lim_{i \rightarrow +\infty }\!\!\frac{-Z_{i-1}^n}{Z_i^p}=0.\!\!\!\!\!
\end{aligned}
\end{equation}
Then, by following the similar approach, we deduce that
 \begin{equation}
\label{A5}
\begin{aligned}
\lim_{i \rightarrow +\infty }\!\frac{w_{2i-1}}{S_i^{p}} \!=\!\! \lim_{i \rightarrow +\infty } \! \frac{T}{\int_{0}^{0.5T}\!\!\sin(\omega r)dr \!\cdot\! e^{f(w_{2i-2})}}\!=\!0,
\end{aligned}
\end{equation}
From (\ref{A4}) and (\ref{A5}), there is
  \begin{equation}
  \label{A6}
\begin{aligned}
\lim_{i \rightarrow +\infty }\frac{w_{2i-1}-\int_{0}^{w_{2i-1}}\mathcal{N}_{n}(\nu)d\nu}{\int_{0}^{w_{2i-1}}\mathcal{N}_p\nu)d\nu}=0.
\end{aligned}
\end{equation}

Secondly, for $\{w_{2i}\}$, we can similarly deduced that
   \begin{equation}
\label{A7}
\begin{aligned}
\lim_{i \rightarrow +\infty }\frac{w_{2i}+\int_{0}^{w_{2i}}\mathcal{N}_{p}(\nu)d\nu}{-\int_{0}^{w_{2i}}\mathcal{N}_n(\nu)d\nu}=0
\end{aligned}
\end{equation}
In view of \eqref{A6} and \eqref{A7}, \eqref{6} holds for $w >0$.

\emph{Case 2($w\!<0$):} Since $f(\nu)$ is an even function, $\mathcal{N}(\nu)$ is an odd function. Then it is clear that \eqref{6} holds for $w<0$.

Combining \emph{Case 1} and \emph{Case 2}, it implies that $\mathcal{N}(\nu) \in \mathcal{N}$.

(2). Let $ w_{i}=\frac{1}{4}iT, i\in \mathbb{N}^{+}$. If two sequences $\{w_{4i+1}\}, \{w_{4i-1}\}$ are chosen, (\ref{6}) can be certified by following the proof process in (1). Therefore, $\mathcal{N}(\nu) \in \mathcal{N}$.

The above analysis confirms (i). Then, when (\ref{31}) is used in controllers \eqref{27}, from Lemma \ref{Lemma 3} and (\ref{30}), it shows that Theorem \ref{theorem 2} holds and control objectives are achieved.
\end{ProofNoIndent}

\end{document}